\documentclass[conference]{IEEEtran}
\IEEEoverridecommandlockouts
\usepackage{cite}
\usepackage{amsmath,amssymb,amsfonts}
\usepackage{algorithmic}
\usepackage{graphicx}
\usepackage{textcomp}
\usepackage{tabularx}
\usepackage{booktabs}
\usepackage{xcolor}
\usepackage{ifthen}
\usepackage{bbm}
\usepackage{enumitem}
\usepackage{siunitx}
\def\BibTeX{{\rm B\kern-.05em{\sc i\kern-.025em b}\kern-.08em
    T\kern-.1667em\lower.7ex\hbox{E}\kern-.125emX}}
\begin{document}

\title{TITAN: DRL-Based Dynamic Scheduling and Resource Allocation for Tactile Internet\thanks{This work has been submitted to the IEEE for possible publication. Copyright may be transferred without notice, after which this version may no longer be accessible.}}

\newboolean{showcomments}
\setboolean{showcomments}{true}
\newcommand{\jk}[1]{\ifthenelse{\boolean{showcomments}} {\textcolor{red}{(JK says: #1)}} {} }
\newcommand{\vg}[1]{\ifthenelse{\boolean{showcomments}} {\textcolor{blue}{(VG says: #1)}} {} }
\newcommand{\jc}[1]{\ifthenelse{\boolean{showcomments}} {\textcolor{red}{(JC says: #1)}} {} }
\newcommand{\qf}[1]{\ifthenelse{\boolean{showcomments}} {\textcolor{pink}{(QF says: #1)}} {} }

\author{
\IEEEauthorblockN{
Qingrui Fan\IEEEauthorrefmark{1}\IEEEauthorrefmark{2},
Vineet Gokhale\IEEEauthorrefmark{1}\IEEEauthorrefmark{2},
Jayakrishnan Nair\IEEEauthorrefmark{3},
Johannes Cottyn\IEEEauthorrefmark{1}\IEEEauthorrefmark{2}
}

\IEEEauthorblockA{
\IEEEauthorrefmark{1}
\textit{Department of Industrial Systems Engineering and Product Design,
Ghent University}, Ghent, Belgium\\
\IEEEauthorrefmark{2}
\textit{Industrial Systems Engineering (ISyE), Flanders Make},
Kortrijk, Belgium\\
\IEEEauthorrefmark{3}
\textit{Department of Electrical Engineering,
Indian Institute of Technology Bombay}, Mumbai, India\\
\{qingrui.fan, vineet.gokhale, johannes.cottyn\}@ugent.be;
jayakrishnan.nair@ee.iitb.ac.in
}
}
\maketitle

\begin{abstract}
Tactile Internet (TI) applications require ultra-reliable, low-latency tactile communication alongside video traffic. However, tactile Quality of Service (QoS) requirements vary with the interaction context, making fixed resource allocation inefficient. In this paper, we propose \textit{TITAN} -- \textit{Two-Timescale Integrated Tactile Internet over 5G Access Networks}, a dynamic scheduling and resource allocation (SRA) scheme. TITAN receives time-varying tactile QoS profiles from a higher-layer module and adapts SRA to the current QoS profile and network state. A deep reinforcement learning-based agent splits the available network bandwidth between tactile and video streams at a slow timescale, while a fast-timescale scheduler allocates resource blocks among active user equipment. Through simulations, we show that TITAN simultaneously satisfies both tactile and video QoS targets and achieves a performance gain of 7.64 percentage points (pp) in the joint satisfaction rate over the best-performing static SRA baseline.
\end{abstract}

% \begin{abstract}
% Real-time Tactile Internet interaction between an operator and a teleoperator involves continuously evolving contact hardness and interaction speed at the application layer, giving rise to diverse, time-varying tactile quality of service (QoS) requirements that must be met through adaptive scheduling and resource allocation (SRA) at the medium access control layer. To this end, we propose \textit{TITAN} (\textit{Two-Timescale Integrated Tactile Internet over 5G Access Networks}), in which a slow-timescale deep reinforcement learning agent dynamically slices bandwidth between the tactile and video streams, while a fast-timescale scheduler allocates resource blocks among the user equipment.
% Simulation results show that TITAN improves the joint tactile-video QoS satisfaction ratio by 7.64 percentage points (\SI{9.62}{\%} relative) over the best-performing static SRA scheme while adaptively meeting both streams' QoS requirements.
% \end{abstract}


\section{Introduction}
\begin{figure}[!t]
    \centering
    \includegraphics[width=0.8\linewidth]{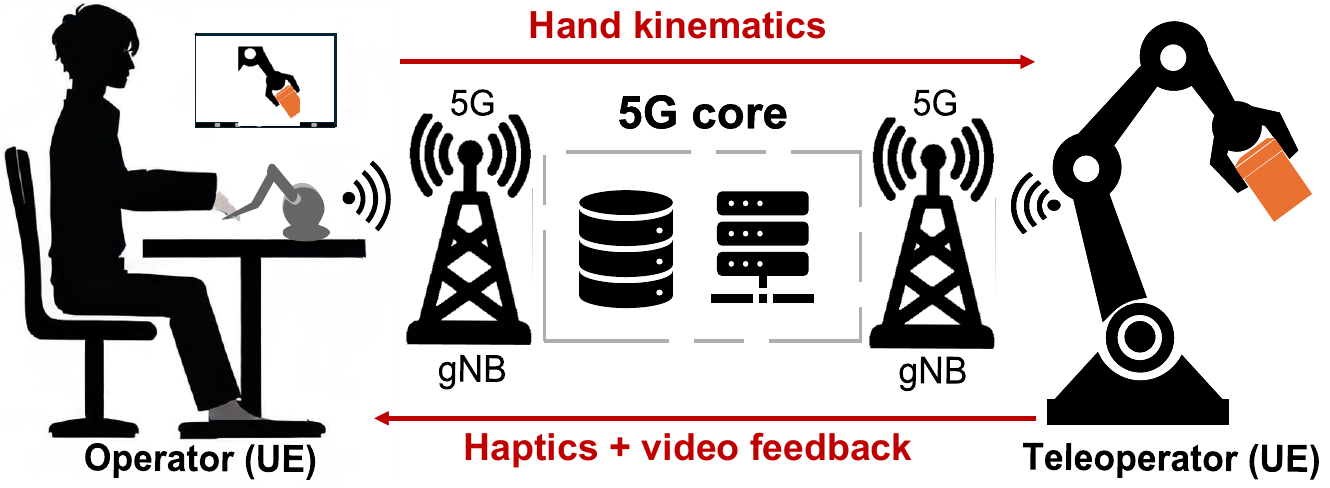}
    \vspace{-5pt}
    \caption{A typical Tactile Internet application running over a shared 5G network showing the operator, remote teleoperator, and the 5G network connecting both.}
    \label{TI_framework}
    \vspace{-10pt}
\end{figure}

% On top of current communication systems that can already offer effective message transmission and video streaming, there is still a distinct gap in communication technologies to facilitate physical interactions across continents. 
While modern communication systems excel at message transmission and video streaming, they remain ill-equipped to support the ultra-reliable, low-latency communication (URLLC) crucial for physical interactions across continents.
The Tactile Internet (TI) aims to bridge this gap, enabling mission- and life-critical applications like remote surgery and industrial teleoperation \cite{fettweis2014tactile}. 
% Driven by the recent advances in  TI, a growing number of emerging applications and use cases such as remote surgery and industrial teleoperation, once considered difficult to envision, are beginning to materialise and continuously evolve. 
As illustrated in Fig.~\ref{TI_framework}, the human operator's kinematics (hand movements) are transmitted over a 5G network to control the teleoperator, which in turn transmits haptic (force feedback) and video feedback to the operator. This heightened experience of immersion enables the operator to manipulate the remote environment with a high level of precision. However, realizing such TI interactions poses challenging Quality of Service (QoS) requirements on the underlying communication networks. The kinematic and haptic signals/streams, jointly referred to as \textit{tactile signals/streams}, demand URLLC.

\textbf{QoS requirements are dynamic:} A large body of TI studies adopts a stringent, uniform QoS target of \SI{1}{ms} latency and over \SI{99}{\%} reliability\footnote{Reliability is defined as the fraction of the generated packets that are delivered within their latency deadlines.} across all TI applications~\cite{mirande2026optimal,ye2026dual}. On the contrary, the TI standardization work recognizes that the QoS requirements are non-uniform and application-dependent~\cite{holland2019}. Furthermore, many prior studies have established that the QoS requirements are dictated by the interaction context. Fast hand movements and interaction with stiff (hard) surfaces generate rapidly varying tactile signals, amplifying the consequences of latency and reliability in terms of position and force errors. Hence, they demand tighter QoS requirements than slow movements and interaction with softer surfaces~\cite{ferrell1966remote,rakita2020effects,du2024sensory,colgate1997passivity,franken2011bilateral,haddadi2009bounded}. In real-world telesurgery, the surgeon interacts with a wide variety of surfaces, such as bones, muscles, and soft tissues, and at varying speeds, all within a single session depending on the medical procedure being conducted. This creates a time-varying QoS requirement within a single TI session. Furthermore, recent teleoperation experiments have shown that the impact of latency on overall performance depends on interaction context and the operators' adaptation to the changes in the remote environment and the network ~\cite{Kroep2025_tmc,kroep2025breaking}. Therefore, the required QoS levels can vary even within a single TI session and assuming a uniform QoS target is unnecessarily rigid.
% To summarise, the QoS requirements for effective teleoperation are non-uniform across TI applications and time-varying within a TI application. 
% context-dependent, with context determined by the remote environment, network behaviour, and the operator's ability to adapt to these changes in real-time. This strongly suggests that QoS requirements are dynamic, showing fluctuations even within the same TI session. 

% the precise QoS requirements vary even within a single application session, as the 
% requiring careful design and deployment of dynamic network algorithms. 
% On the contrary, a vast majority of the TI literature treats the most stringent \SI{1}{ms} latency and \SI{99.999}{\%} as the common QoS requirement without taking the application-level dynamics into account.    

\textbf{Need for QoS-aware resource management}: Existing TI-specific schemes for scheduling and resource allocation (SRA) commonly provision resources for fixed, worst-case QoS requirements, such as \SI{1}{ms} latency and over \SI{99}{\%} reliability~\cite{she2016cross, mirande2026optimal, kokkinis2025deep, ye2026dual}. 
% network designs consider the static QoS requirement (\SI{1}{ms} and \SI{99.999}{\%}) as the holy grail. 
This static treatment overlooks the interaction-dependent QoS requirements described above and overprovisions resources for TI traffic. 
% has led to several designs proposing aggressive provisioning of 5G network resources to TI streams. 
This comes at the expense of non-TI network traffic, potentially disrupting the associated services. Such sub-optimal schemes can cause severe underperformance of the network and limit the network's capacity to support concurrent applications. 
% thereby degrading the performance of non-TI applications. 
The problem is exacerbated under network congestion. Furthermore, maintaining unnecessarily rigid performance guarantees requires faster coordination between the TI user equipment (UE) and gNodeB (gNB) (see Fig.~\ref{TI_framework}). This involves frequent state exchanges, which introduce signaling overhead and latency and consume additional resources~\cite{maghsoudnia2024ultra}. For graceful coexistence between TI and other applications, dynamic SRA schemes that allocate resources to match the current QoS requirements are essential.

Motivated by the need for dynamic SRA for TI applications, 
we propose \textit{TITAN} - Two-Timescale Integrated Tactile Internet over 5G Access Networks. Focusing on the uplink, TITAN adopts a cross-layer approach in which the radio access network (RAN) module adapts its transmission decisions to time-varying QoS requirements supplied by a higher-layer interaction-aware module. It is important to note that the focus of this work is purely on the design of the adaptive SRA strategy based on the supplied dynamic QoS requirements and not on estimating the QoS requirements. The primary contributions are as follows:
\begin{itemize}[leftmargin=*]
\item Learning from existing works, we classify the TI requirements into four different QoS classes based on the surface properties and the operator's motion dynamics.
\item We propose TITAN: a realistic, dynamic two-timescale SRA strategy that reduces UE-gNB control-signaling overhead. On the slow timescale, we formulate bandwidth slicing as a deep reinforcement learning (DRL) problem and employ Proximal Policy Optimization (PPO) to dynamically allocate resources between tactile and video streams. On the fast timescale, we schedule UEs on 5G resource blocks (RBs).
\item We introduce two metrics for evaluating SRA schemes designed for TI applications.  
\item Through extensive simulations, we demonstrate that TITAN achieves an absolute improvement of \SI{7.64}{pp} in joint tactile-video QoS satisfaction over the best-performing static SRA scheme.
\end{itemize}

Subsequently, we review the relevant literature on TI QoS and SRA schemes in Section II and present the TITAN scheme in Section III. We conduct TITAN's performance evaluation in Section~IV and present our conclusions in Section~V.

\section{Related Literature}
\subsection{TI's Dynamic QoS Requirements}
\label{subsec:dynamicQoS}
The impact of communication latency on bilateral teleoperation has been widely studied in terms of position and force errors, stability, and perceived task performance~\cite{ferrell1966remote,rakita2020effects,du2024sensory,colgate1997passivity,franken2011bilateral,haddadi2009bounded}. In particular, fast operator movements are more sensitive to latency than slow movements because position errors increase with motion velocity~\cite{ferrell1966remote,rakita2020effects,du2024sensory}. Similarly, interactions involving hard contact surfaces can amplify force errors and stability constraints, whereas soft surfaces are typically more tolerant to latency and force
distortion~\cite{colgate1997passivity,franken2011bilateral,haddadi2009bounded}.

Recent studies further indicate that the communication performance required for effective teleoperation is interaction dependent rather than fixed throughout a session. Operator intent, task phase, remote-environment characteristics, and the adopted teleoperation-control mechanism can affect the latency that can be
tolerated while preserving task performance~\cite{Kroep2025_tmc,kroep2025breaking}. Consequently, a single teleoperation session can alternate between demanding phases, such as fast motion or hard contact, and more delay-tolerant phases involving slow motion and/or soft contact. 

In summary, teleoperation studies show that tactile QoS is not a one-size-fits-all entity; it keeps changing within a single TI session and across TI sessions.
% rich literature exists showing that hard contact surfaces generate higher force feedback, requiring that it be delivered faster compared to soft surfaces. 
% To summarize, fast operator movements and hard objects pose tighter QoS requirements than slower movements and soft objects. Furthermore, within the same TI session, the operator could exhibit different movement speeds and interact with varying contact surfaces. 
% Hence, any SRA approach targeting static QoS requirements cannot be directly applied to TI applications, as it considers .

\subsection{5G SRA for Tactile Internet}
Although 5G URLLC has attracted extensive research attention \cite{haque2025comprehensive}, only a limited number of studies have investigated SRA for TI traffic.~\cite{she2016cross} employed a transmission optimization strategy,
considering queueing delay, transmission delay, and packet dropping to meet stringent static QoS requirements.
Existing approaches include optimization-based schemes, such as Lyapunov optimization\cite{golmohammadi2024lyapunov}, Lagrangian water-filling \cite{mirande2026optimal}, and DRL-based approaches using soft actor-critic (SAC)~\cite{kokkinis2025deep} and deep deterministic policy gradient (DDPG) \cite{ye2026dual}. While the above works demonstrate the value of
cross-layer optimization and learning-based adaptation, they retain
static QoS assumptions and do not account for QoS variations during a teleoperation session. 

Among the above works, \cite{kokkinis2025deep} is most closely related to our work because it adopts a two-timescale design in which bandwidth slicing is separated from fine-grained scheduling. However, they consider static QoS requirements, and their SRA scheme requires frequent UE-gNB state exchanges and fine-grained scheduling updates, thereby increasing control signaling overhead in latency-sensitive TI systems. 
In contrast, TITAN incorporates (i) time-varying tactile QoS requirements supplied by a higher-layer interaction-aware module and (ii) fine-grained scheduling aimed at reducing the frequency of control updates while also coping with the dynamic QoS requirements.

\section{Architecture of TITAN}
\subsection{Defining Tactile QoS Classes}\label{haptic class}
As discussed in Sec.~\ref{subsec:dynamicQoS}, we consider TI interactions involving varying operator speeds and contact-surface stiffness. To this end, we heuristically define the following four tactile QoS classes along with their respective QoS requirements:
% motivated by the fact that delay-induced position and force errors increase with operator velocity and contact stiffness:
\begin{itemize}[leftmargin=*]
    \item \textit{Hard-fast}: Interaction with a hard surface at high operator velocity, with a QoS latency requirement of \SI{1}{ms}. 
    \item \textit{Soft-fast}: Interaction with a soft surface at high operator velocity, with a QoS latency requirement of \SI{4}{ms}.
    \item \textit{Hard-slow}: Interaction with a hard surface at low operator velocity, with a QoS latency requirement of \SI{7}{ms}.
    \item \textit{Soft-slow}: Interaction with a soft surface at low operator velocity, with a QoS latency requirement of \SI{10}{ms}.
\end{itemize}
For ease of illustration, the reliability requirement for all tactile QoS classes is set to \SI{99}{\%}. We consider a video latency requirement of \SI{50}{ms} and a reliability requirement of \SI{90}{\%}~\cite{holland2019}.

\subsection{Schematic Framework of TITAN}
\begin{figure}[!t]
    \centering
    \includegraphics[width=0.9\linewidth]{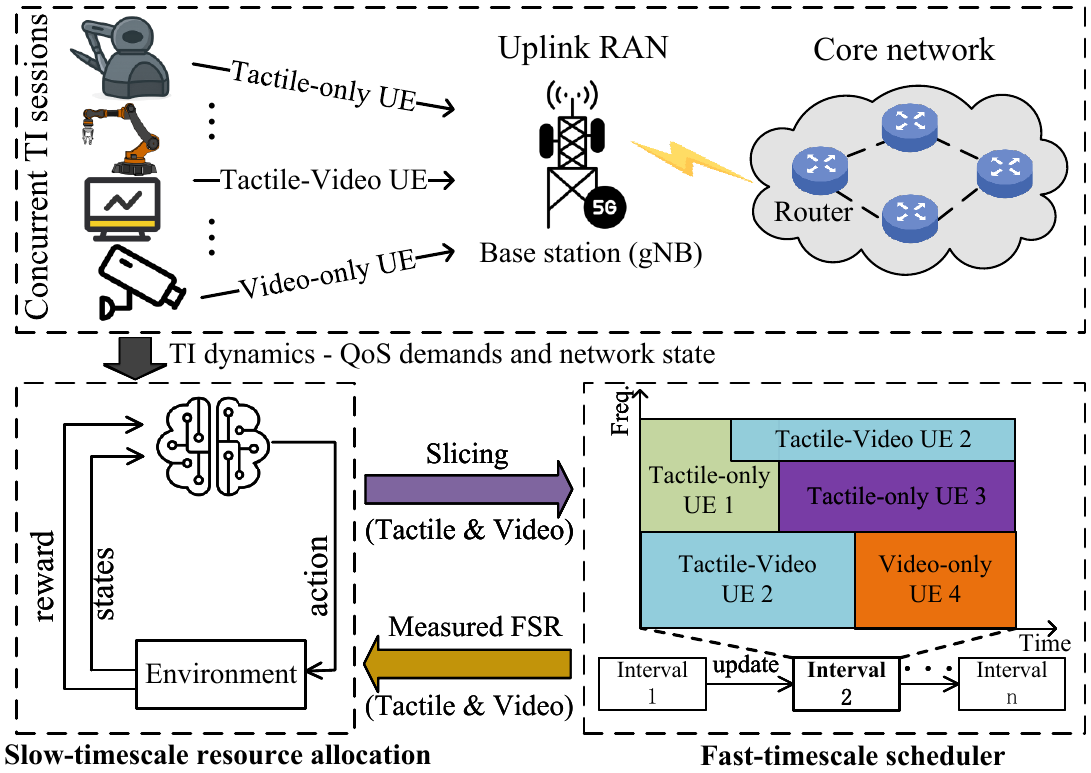}
    \vspace{-7pt}
    \caption{Framework of the proposed TITAN scheme showing slow-timescale resource allocation (bandwidth slicing) and fast-timescale scheduling of 5G resource blocks to TI and non-TI UEs.}
    \vspace{-10pt}
    \label{scheme diagram}
\end{figure}

% \jk{Highlight that we are only considering uplink communications?}
As illustrated in Fig.~\ref{scheme diagram}, we consider a generic uplink RAN where multiple TI and non-TI UEs contend for a common pool of RBs. Each UE belongs to one of the following categories based on the traffic they generate: (i) tactile-only UEs (operators generating kinematic traffic), (ii) tactile-video UEs (teleoperators generating both haptic and video traffic), and (iii) video-only UEs (representing non-TI traffic coexisting in the network). As stated earlier, we assume that a higher-layer interaction-aware module at each UE provides the time-varying QoS demands to the gNB, which also gathers the network-state information such as active UEs. 
% , are exposed to the medium access control (MAC) layer and shape the resource allocation strategy accordingly.

\textbf{Two-timescale design:} At the data link layer, SRA operates on two timescales. On the slow timescale, a trained DRL agent allocates radio resources, i.e., dynamically slices the available bandwidth (RBs) between tactile and video streams every $T_{DRL}$ time units based on the observed state. On the fast timescale, a scheduler updates UE-specific RB reservations every $T_s$ time units. Specifically, the scheduler employs an urgency-ordered round-robin rule for the tactile stream and a least-laxity-first rule for the video stream. It is worth noting that the SRA computation happens at the gNB. The UEs comply with the SRA schedule published by the gNB.
% This slow-timescale design reduces the computational burden of DRL training by avoiding TTI-level decisions while adapting bandwidth slicing to evolving aggregate service demands. Meanwhile, the fast-timescale frame-level scheduler avoids frequent TTI-level rescheduling, thereby reducing scheduling overhead and associated control-signaling demands.

\subsection{Slow-timescale DRL-based Resource Allocation}
While online optimization can be computationally prohibitive, DRL enables lightweight inference by shifting the computational burden to offline training. To minimize transmission time interval (TTI)-level training complexity, we propose a slow-timescale DRL framework, leveraging the fact that tactile interaction states, namely contact stiffness and motion velocity, evolve over periods longer than a TTI.
% Repeatedly solving complex optimization problems under dynamic network conditions can be computationally prohibitive. DRL shifts much of the computational burden to offline training, enabling lightweight online inference. 
% However, transmission time interval (TTI)-level DRL training remains computationally demanding. 
% Since tactile interaction states, characterised by contact stiffness and motion velocity, typically evolve over much longer timescales than the TTI, slow-timescale DRL can reduce training complexity while retaining adaptability to changing interaction conditions.
We adopt PPO~\cite{schulman2017proximal} to learn a stochastic policy over discrete RB allocations. Its clipped surrogate objective discourages large policy updates, promoting training stability under stochastic traffic, channel, and QoS variations. Unlike continuous action algorithms, such as DDPG~\cite{ye2026dual} and SAC~\cite{kokkinis2025deep}, that require their continuous outputs to be mapped to discrete RB allocations, or value-based methods, such as deep Q-networks, lacking direct policy optimizations, PPO directly optimizes the discrete allocation policy. The following details our PPO formulation focusing on the agent, action, state, and reward definitions.
% This subsection details our PPO-based formulation for bandwidth slicing, describing the agent, action space, state space, and reward function.

\subsubsection{Agent}
This is a deep neural network residing at the gNB. The DRL agent observes the state, selects an action accordingly, and receives a reward that guides the update of the policy network's weights during training. Once trained offline, the learned policy is deployed for online execution.

\subsubsection{Action}
% In the proposed formulation, the agent's action specifies the bandwidth slicing between haptic and video services. For example, with a total of 15 RBs, an action allocating 5 RBs to haptic service implicitly assigns the remaining 10 to video service. 
At each decision epoch $k$, i.e., the beginning of the $k$-th DRL decision interval, the agent determines the bandwidth slicing between the tactile and video streams. 
% For example, with a total of 21 RBs, allocating 8 RBs to the tactile stream implicitly assigns the remaining 13 RBs to the video stream.
% This slicing is then held fixed for the subsequent DRL update interval.

\subsubsection{State}
At each decision epoch $k$, the agent observes a vector $\mathbf{s}_k$ comprising (i) active UE counts by category and tactile QoS class composition (how many UEs belong currently to each class); (ii) the action selected at epoch $k-1$; (iii)
the tactile and video frame success rates (FSRs) measured over interval \(k-1\), as defined below; and (iv) the signal-to-noise ratio (SNR) of the active UEs. All of these components are causally available at the decision boundary, and no future information, such as traffic arrivals or UE activity states, is available to the agent.

\subsubsection{Reward}
At the end of interval $k$, the environment returns a scalar reward that jointly promotes satisfaction of both tactile and video QoS requirements under time-varying conditions.
Let $\mathcal{C} = \left\{ \text{hard-fast}, \text{soft-fast}, \text{hard-slow}, \text{soft-slow} \right\}$ denote the set of four tactile QoS classes defined in Sec. \ref{haptic class}.
Let $B_{c,k}^{h}$, $A_{c,k}^{h}$, and $U_{c,k}^{h}$ denote, respectively, the backlog of class-$c$ packets carried into interval $k$, the new class-$c$ arrivals during interval $k$, and the class-$c$ packets remaining unresolved at the end of interval $k$. A packet is considered resolved once it is either successfully delivered or discarded due to a QoS violation. The resolved packet count is then defined as
\[
D_{c,k}^{h}=B_{c,k}^{h}+A_{c,k}^{h}-U_{c,k}^{h},
\]
i.e., unresolved packets are excluded from $D_{c,k}^{h}$ and accounted for only when resolved in a subsequent interval. The active tactile QoS class set for interval $k$ is defined as $\mathcal{C}_k=\{c\in\mathcal{C}:D_{c,k}^{h}>0\}\subseteq\mathcal{C}$. For each class $c\in\mathcal{C}_k$, the tactile FSR is
\begin{equation}
s_{c,k}^{h} = \frac{N_{c,k}^{\mathrm{h,on}}}{D_{c,k}^{h}},
\end{equation}
where $N_{c,k}^{\mathrm{h,on}}$ denotes the number of packets delivered within their respective QoS deadlines. The maximum normalized tactile QoS shortfall is then
\begin{equation}
\delta_k^{h} = \max_{c\in\mathcal{C}_k} \frac{\left[\eta_h-s_{c,k}^{h}\right]^+}{\eta_h},
\end{equation}
where $[x]^+\triangleq\max\{x,0\}$ and $\eta_h$ is the target tactile FSR, set to \SI{99}{\%}; by convention, $\delta_k^h\triangleq 0$ when $\mathcal{C}_k=\emptyset$.

The resolved frame count of the video stream is defined analogously,
\begin{equation}
D_k^{v} = B_k^v + A_k^v - U_k^v,
\end{equation}
where $B_k^v$ is the backlog of video frames carried into interval $k$, $A_k^v$ is the number of new frame arrivals during the interval, and $U_k^v$ is the number of frames not yet resolved by the end of the interval. The video FSR is defined as
\begin{equation}
s_k^{v} = \frac{N_k^{\mathrm{v,on}}}{D_k^{v}},
\end{equation}
where $N_k^{\mathrm{v,on}}$ denotes the number of video frames completely delivered within the defined deadline. The normalized video QoS shortfall is defined as
\begin{equation}
\delta_k^v = \frac{\left[\eta_v-s_{k}^{v}\right]^+}{\eta_v},
\end{equation}
where $\eta_v$ is the target video FSR, set to \SI{90}{\%}; by convention, $\delta_k^v \triangleq 0$ when $D_k^v=0$.

To prevent satisfactory performance in one stream from compensating for a QoS violation in the other, the joint normalized shortfall is defined as $\delta_k^{J} = \max\{\delta_k^{h},\delta_k^{v}\}$.
% \begin{equation}
% \end{equation}
The overall reward for decision interval $k$ is then
\begin{align}
\mathcal{R}_k = b_J \mathcal{I}_k^{J} - \lambda_J \delta_k^{J},
\end{align}
where $\mathcal{I}_k^J = \mathbf{1}\!\left\{\mathcal{C}_k \neq \emptyset,\,D_k^v>0,\,\delta_k^h=0,\,\delta_k^v=0\right\}$ is the joint satisfaction indicator, equal to 1 if and only if both tactile and video streams are active during interval $k$ and each stream meets its respective QoS target, and 0 otherwise. $b_J = 1$ and $\lambda_J = 1$ are fixed weights scaling the joint satisfaction bonus and joint shortfall penalty, respectively. The continuous shortfall penalty supplies a dense learning signal proportional to the severity of the dominant QoS violation, facilitating smoother learning.

\subsection{Fast-timescale Scheduling}
Once the DRL agent determines the tactile and video slices (RBs), the fast-timescale scheduler allocates these RBs among the currently active UEs. Conventional scheduling based on TTI-level signaling includes frequent queue-status reporting, scheduling-request processing, and grant signaling. With a short TTI of \SI{0.25}{ms}, this results in significant overhead and latency for TI’s applications. To alleviate this, the fast-timescale scheduler constructs UE-specific RB reservations only at the beginning of each scheduling interval $T_s$ and keeps them fixed throughout the interval, thereby avoiding additional control signal exchanges.

\subsubsection{Tactile Frame Scheduling}
Due to its small size, each tactile frame is conventionally transmitted in a separate packet~\cite{holland2019}. TITAN schedules tactile packets in two stages. First, the scheduler processes the backlog packets based on their QoS targets and current SNR. Packets that cannot be delivered before their deadlines are proactively discarded and counted as QoS violations. The remaining packets are assigned the RBs at their latest feasible transmission opportunities within the scheduling interval. This strategy preserves capacity for packets with potentially more urgent requirements.
Second, the remaining RBs are reserved for expected tactile packet arrivals during the upcoming interval across all active UEs, including those with empty backlogs. 
% tactile packets expected to arrive during the forthcoming scheduling interval. This stage considers every tactile UE active at the scheduling boundary, including UEs with empty backlogs. 
The scheduler estimates packet arrivals and calculates RB budgets based on the UE's packet generation characteristics and SNR. The UE-associated RB bundles are then distributed via an urgency-ordered round-robin procedure.
At runtime, these reserved RBs dynamically serve the UE's head-of-line pending packet, whether backlogged or newly arrived.
% packet or a newly arrived packet of the same UE, depending on which packet is pending when the RB becomes available. 
% This UE-level reservation improves resource utilisation while avoiding packet-level rescheduling and additional request-grant exchanges. 
% Backlog and predicted packets can thus implicitly substitute for each other within a UE, improving utilisation without packet-level rescheduling or additional request-grant exchanges.

% \begin{figure}
%     \centering
%     \includegraphics[width=0.8\linewidth]{frame_level_scheduler.pdf}
%     \caption{Fast-timescale scheduler diagram. 
%     {\color{red}assist for review of this subsection}}
%     \label{frame_level_scheduler}
% \end{figure}
\begin{figure}[!t]
    \centering
    \includegraphics[width=0.8\linewidth]{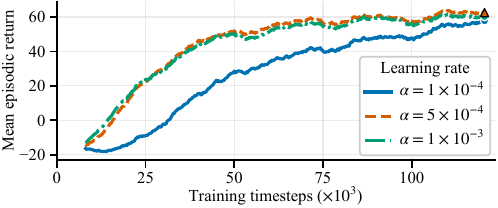}
    \vspace{-12pt}
    \caption{PPO training convergence under three learning rates, showing the 100-episode moving average of the episodic return over training timesteps.}
    \label{PPO_training}
    \vspace{-10pt}
\end{figure}

\subsubsection{Video Frame Scheduling}
Unlike tactile traffic, the arrival times of video frames are largely predictable. At each scheduling interval boundary, the scheduler jointly considers the existing video backlog and the expected arrivals during the subsequent scheduling interval. We use the average frame-size estimate for RB allocations. Any mismatch between the estimated and actual frame sizes only affects the current scheduling interval and is corrected at the next scheduling boundary.
Frames are prioritized by a least-laxity-first rule based on their deadlines, earliest service times, remaining or predicted bits, and UE-specific transmission rates, subject to per-UE first-in, first-out ordering. Selected frames are scheduled non-preemptively across a fixed RB reservation grid. Partially transmitted but feasible frames carry their remaining bits into the next interval, while infeasible frames are discarded.
% \vspace{-4pt}

\section{Performance Evaluation of TITAN}
In this section, we present the detailed performance setup and evaluation of TITAN.
% \vspace{-13pt}
\begin{figure*}[!t]
    \centering
    \includegraphics[width=0.95\linewidth]{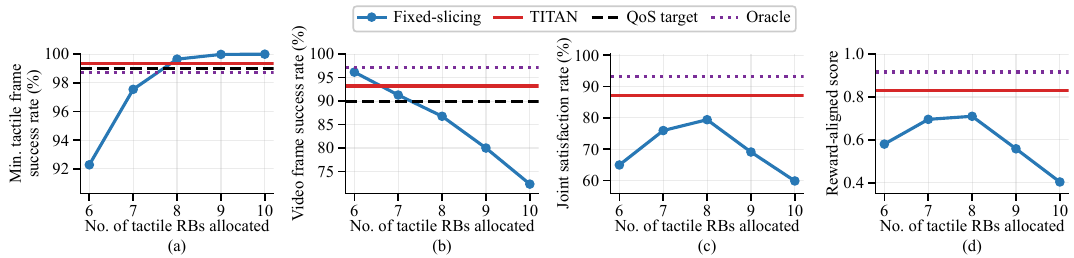}
    \vspace{-12pt}
    \caption{Performance comparisons among fixed-slicing, TITAN, and the Oracle reference. (a) Minimum tactile Frame Success Rate (FSR) across QoS classes, (b) Video FSR, (c) joint tactile-video performance measured by Joint Satisfaction Rate (JSR), and (d) joint tactile-video performance measured by Reward-Aligned Score (RAS).} 
    \vspace{-10pt}
    % \vg{remove the numbers you show on the plots, and also the text "TITAN = ..." and "Oracle  = ..."}
    % \qf{unify the font size, font style}
    \label{combined_figure}
\end{figure*}

\subsection{Performance Comparison}
Direct numerical performance comparison of TITAN with \cite{kokkinis2025deep} is not included here because their system model, traffic assumptions, and QoS formulation differ from the time-varying QoS profile considered here. We instead compare TITAN against the following schemes:
% \begin{enumerate}[leftmargin=*]
%     \item Best fixed-slicing: 
%     \item Oracle: 
% \end{enumerate}
\begin{itemize}[leftmargin=*, nosep]
    \item \textit{Fixed-slicing}: This is a static SRA scheme where the RBs are split over tactile and video streams in a QoS- and network-agnostic manner.
    \item \textit{Oracle}: This is an ideal, noncausal scheme used for computing the upper bound of performance. Oracle uses hindsight knowledge of actual traffic arrivals, channel conditions, and service outcomes in the subsequent decision interval to exhaustively evaluate the optimal RB split for achieving the best joint QoS performance in each DRL decision interval. While this scheme is non-realistic, we use it to assess TITAN's performance against the upper bound.
\end{itemize}

\subsection{Performance Metrics}
% Each scheme is evaluated over $R=35$ independent realizations of \SI{40}{s} each, comprising $K=80$ decision intervals of \SI{500}{ms} per realization ($RK=2800$ intervals in total). For interval $(r,k)$, let $\delta_{r,k}^h$ and $\delta_{r,k}^v$ denote the normalized tactile and video QoS shortfalls, respectively. Assuming that both services have resolved traffic in every
% evaluation interval, the joint satisfaction indicator is defined as $I_{r,k}^{\mathrm{joint}}=\mathbf{1}(\delta_{r,k}^h=0 \text{ and } \delta_{r,k}^v=0)$. The joint satisfaction ratio (JSR), which also reflects the scheme's adaptability to time-varying conditions, is given by
% K=80, R=35, T_DRL=500ms, T=40s. T_s = 50ms
% We now introduce the two performance metrics we devised for evaluating the fine-grained performance of TITAN.
Denoting the duration of each simulation run as $T$, the total number of DRL decision intervals in a simulation run is given by $K=T/T_{DRL}$ . Each aforementioned scheme is evaluated over $R$ independent simulation runs, yielding a total of $RK$ DRL decision intervals per scheme.
For each DRL decision interval $k$ in simulation run $r$, let $\delta_{r,k}^h$ and $\delta_{r,k}^v$ denote the normalized tactile and video QoS shortfalls, respectively. 

\subsubsection{Joint Satisfaction Rate (JSR)}
We define JSR as the fraction of DRL decision intervals in which both tactile and video QoS targets are simultaneously met. JSR is given by
% \qf{we are not using the same indicator in 2025 paper, they defined the SR as the the proportion of simulation time during which the latency, packet-loss, and data-rate requirements of both operator and teleoperator are satisfied}
\begin{equation}
\mathrm{JSR}
=
\frac{\sum_{r=1}^{R}\sum_{k=1}^{K} \, I_{r,k}^{\mathrm{joint}}}
{RK},
\end{equation}
where $I_{r,k}^{\mathrm{joint}}
=
\mathbf{1}\!\left(
\delta_{r,k}^{h}=0 \;\land\; \delta_{r,k}^{v}=0
\right)$
% \begin{equation}   
% \end{equation}
is a joint-satisfaction indicator that equals one only when both tactile and video streams satisfy their respective QoS targets during the interval $(r,k)$. Therefore, JSR directly measures an SRA scheme's ability to jointly satisfy both streams' QoS requirements under time-varying conditions, rather than evaluating each stream independently. However, JSR does not distinguish between intervals with different magnitudes of QoS violations.

\subsubsection{Reward-Aligned Score (RAS)} To overcome JSR's limitation, we devise RAS. Let us define a \textit{per-interval score} as $s_{r,k}
=
I_{r,k}^{\mathrm{joint}}
-
\max\{
\delta_{r,k}^{h},\,
\delta_{r,k}^{v}
\}$,
% \begin{equation}
% \end{equation}
which rewards joint QoS satisfaction while penalizing the dominant normalized
QoS shortfall. 
Based on this, we define RAS as the average $s_{r,k}$ over all simulation runs and DRL decision intervals:
\begin{equation}
\mathrm{RAS}
=
\frac{\sum_{r=1}^{R}\sum_{k=1}^{K} \, s_{r,k}}
{RK}.
\end{equation}
Unlike the percentage-based JSR, RAS lies in $[-1,1]$ and captures both the frequency and severity of QoS violations. 

In addition to JSR and RAS, we separately report tactile and video FSRs as their respective reliability metrics for all considered schemes.
% \vg{Link FSR to reliability.}

\subsection{Network Traffic Generation}
To mimic a realistic, dynamic network scenario, we randomly pick the number of tactile-only, tactile-video, and video-only UEs every \SI{2}{s} from the ranges [4,15], [1,2], and [1,3], respectively.
% For ease of explanation, we use \textit{tactile-bearing UEs} to jointly refer to tactile-only and tactile-video UEs and \textit{video-bearing UEs} to jointly refer to tactile-video and video-only UEs. 
Each tactile stream is independently and equiprobably assigned one of four tactile QoS classes defined in Sec.~\ref{haptic class}.
Motivated by empirical studies, the tactile packet arrivals follow a generalized Pareto distribution \cite{mondal2022global} with a packet size of \SI{48}{\byte} and a shifted-exponential inter-arrival process (mean of \SI{2}{ms}), resulting in an average packet rate of \SI{500}{\text{packets}/\second}.
Video frames arrive at \SI{30}{\text{fps}} (independently phased), and their sizes follow a truncated Gaussian distribution~\cite{gapeyenko2024overview}
% $\mathcal{TN}(\mu,\sigma^2,a,b)$~\cite{gapeyenko2024overview}, 
with a mean of \SI{20.83}{\kilo\byte}, standard deviation of \SI{2.18}{\kilo\byte}, and lower and upper bounds of \SI{10.41}{\kilo\byte} and \SI{31.25}{\kilo\byte}, respectively. This results in a mean video load of \SI{5}{Mbps} per video-generating UE (tactile-video and video-only UEs). 
% The frame size follows a truncated Gaussian distribution $\mathcal{TN}(20833, 2188^2, 10417, 31250)$ bytes~\cite{gapeyenko2024overview}. \vg{the squared is not making sense here. Why not express everything in terms of kB?}
% The frame size 

\subsection{TITAN and 5G New Radio (NR) Parameters}
% \vg{Mention values of .}
In this work, we empirically set $T_{DRL}=\SI{500}{ms}$ and $T_s=\SI{50}{ms}$. Let $h$ and $v$ denote the numbers of RBs allocated to tactile and video streams, respectively. 
As per \cite{maghsoudnia2024ultra}, we configure  a total of 21 RBs, i.e. $h+v=21$. Each RB spans 12 subcarriers with a \SI{60}{kHz} subcarrier spacing. We adopt the corresponding \SI{0.25}{ms} NR slot as the TTI  to support millisecond-level QoS requirements. To capture heterogeneous channel conditions, each UE's SNR is independently and uniformly sampled from the range $[6,14]$~dB at the start of each simulation run and held quasi-static thereafter.

\subsection{PPO Training Stage}
In Fig.~\ref{PPO_training}, we present the 100-episode moving average of the PPO agent’s training return under three learning rates. Training with $\alpha = 5\times 10^{-4}$ or $1\times 10^{-3}$ converges significantly faster than training with $\alpha = 1\times 10^{-4}$ as larger learning rates yield larger policy updates. All curves eventually plateau with residual fluctuations from ongoing exploration. Since $\alpha = 5\times 10^{-4}$ yields this fast convergence while reaching a slightly higher and more stable final return, its checkpoint (triangle marker) is selected for subsequent evaluation. The remaining hyperparameters are set as follows: discount factor $\gamma=0.99$, generalized advantage estimation parameter $\lambda_\textit{GAE}=0.95$, clip range $\epsilon=0.2$, rollout length and batch size of 512, 10 optimization epochs per update, entropy coefficient 0.01, and $128\times128$ policy/value networks.

% \begin{figure}
%     \centering
%     \includegraphics[width=0.75\linewidth]{training_curve.pdf}
%     \caption{Training curve of the PPO agent under three learning rates $\alpha$.}
%     \label{training_curve}
% \end{figure}

\subsection{Simulation Results}

\begin{figure}[!t]
    \centering    
    \includegraphics[width=0.9\linewidth]{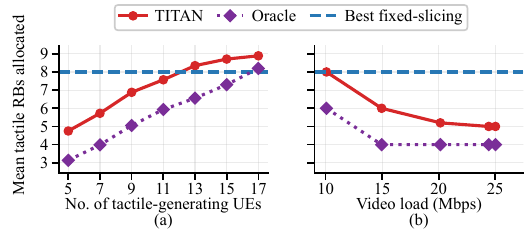}
    \vspace{-11pt}
    \caption{Mean tactile RB allocations of TITAN, best fixed-slicing, and Oracle under varying (a) no. of tactile-generating UEs and (b) video load.}
    \label{Adaptation}
    \vspace{-12pt}
\end{figure}

% \begin{figure}[!t]
%     \centering
%     \includegraphics[width=0.75\linewidth]{Performance_comparison.pdf}
%     \caption{Fixed-slicing trade-off and comparison with the proposed scheme.}
%     \label{Performance_comparison}
% \end{figure}

% \subsubsection{Performance Evaluation}
Each simulation run lasts $T=\SI{40}{s}$, and results are
averaged over $R=35$ independent runs. We first evaluate the sensitivity of
fixed slicing to the tactile RB allocation $h$, sweeping
$h\in\{6,\ldots,10\}$, with the remaining $v=21-h$ RBs assigned to the video stream.
In Fig.~\ref{combined_figure}(a), we report the minimum tactile FSR across all tactile QoS classes. As expected, increasing $h$ improves tactile performance by allocating more RBs to the tactile stream.
However, this improvement comes at the expense of video performance, as shown in
Fig.~\ref{combined_figure}(b), because fewer RBs remain for the video stream.
% In this work, we set the simulation parameters $T=\SI{40}{s}$ and $R=35$.
% We begin by sweeping $h$ over the range [6, 10] for the fixed-slicing scheme. In Fig.~\ref{combined_figure}(a), we plot the minimum tactile FSR measured across the tactile QoS classes. 
% As expected a higher $h$ gives more RBs to tactile streams, thereby improving the tactile FSR as seen in Fig.~\ref{combined_figure}(a). However, this comes at the expense of video FSR as seen in  Fig.~\ref{combined_figure}(b).
Among the fixed-slicing candidates, $h=8$ provides the best overall
trade-off for the considered scenario: it maintains the tactile FSR close to
the QoS target while achieving higher video FSR than configurations with
larger tactile allocations. We therefore use $h=8$ as the
best fixed-slicing baseline.
We also report the FSR performances of TITAN and the Oracle. Unlike the fixed-slicing scheme, TITAN and the Oracle adapt $h$ dynamically rather than depending on a fixed $h$. The Oracle achieves a higher video FSR than TITAN, but at the expense of tactile QoS violations.
TITAN is the only causal scheme that simultaneously satisfies the tactile and
video FSR targets.

% For the chosen experimental  settings, $h=8$ performs the best out of all fixed-slicing configurations.
% For the sake of illustration, we plot the performance of TITAN and Oracle alongside. Firstly, these two schemes are unaffected by the fixed $h$, which is a characteristic of the fixed-slicing scheme. While Oracle's video FSR is higher than that of TITAN, it achieves it at the expense of tactile QoS violations. On the other hand, TITAN clearly outperforms other schemes by simultaneously satisfying both FSR targets. 

In Figs.~\ref{combined_figure}(c) and \ref{combined_figure}(d), we further compare the schemes using JSR and RAS. TITAN achieves a JSR of \SI{87.07}{\%} and an RAS of $0.8286$, compared with \SI{79.43}{\%} and $0.7103$, respectively, for the best fixed-slicing baseline, representing absolute improvements of \SI{7.64}{pp} and $0.1183$, respectively.
The Oracle
achieves the highest JSR and RAS, at \SI{93.36}{\%} and $0.9166$, respectively;
however, it is a noncausal upper bound that relies on future information and
cannot be implemented online. The remaining gap therefore indicates the
potential for prediction-aware SRA, rather than a directly attainable performance target.

We now move to demonstrating the schemes' adaptability to changes in tactile and video streams. First, we fix the number of video-generating UEs (tactile-video and video-only UEs) to three and vary the number of active tactile-generating UEs (tactile-video and tactile-only UEs). As shown in Fig.~\ref{Adaptation}(a), the mean tactile RB allocation increases with the number of tactile-generating UEs, demonstrating that TITAN allocates additional resources to the tactile stream when the observed tactile demand increases. Next, we fix the tactile stream composition to one soft-fast UE and two UEs from each of the remaining tactile QoS classes, and vary the video load. As shown in Fig.~\ref{Adaptation}(b), TITAN
reduces the mean tactile RB allocation as the video load increases, thereby releasing resources for the video stream. Together, these complementary trends show that TITAN adapts the tactile-video resource split to the current stream conditions, unlike a fixed-slicing scheme whose allocation remains constant.

% Alongside, we also show the Oracle's optimal allocation. Fig.~\ref{Adaptation}(a) demonstrates how TITAN adapts RB slicing to observed traffic conditions. As can be seen, the mean tactile RBs increases to approximately 8.9 to keep up with 17 tactile-bearing UEs. In the next set of experiments, we fix the number of tactile-only UEs and evaluate TITAN's adaptation to video load. Specifically, we simulate one soft-fast UEs, and two UEs for each of remaining QoS classes. We vary the video load between \SI{10}{Mbps} to \SI{25}{Mbps}. With increasing video load, we see, in Fig.~\ref{Adaptation}(b), that the RBs allocated to tactile streams reduces implying that TITAN provisions more resources to the video streams. Together, these conditional trends illustrate TITAN's adaptive response to competing tactile and video demands, in contrast to the static allocation of the fixed-slicing scheme.

% , indicating that TITAN allocates fewer tactile RBs under light tactile demand and more under heavy demand.
% Similarly, Fig.~\ref{Adaptation}(b) conditions on a tactile QoS composition of two hard-fast, one soft-fast, two hard-slow, and two soft-slow UEs. As recent video load increases from approximately \SI{10}{Mbps} to \SI{25}{Mbps}, the mean tactile allocation decreases from 8 to approximately 5 RBs, increasing the video allocation from 13 to approximately 16 RBs. 

% \vspace{-5pt}
\section{Conclusion}
This work highlights the limitations of provisioning network resources to TI applications according to static QoS requirements when the requirements of  teleoperation vary over time. TITAN addresses this mismatch through a dynamic SRA scheme for tactile-video streams that uses QoS profiles supplied by a higher-layer module. It combines slow-timescale DRL-based bandwidth slicing between tactile and video streams with fast-timescale UE scheduling over the allocated resources.
% In a first-of-its-kind work, we shed light on the negative implications of assuming static QoS requirements for TI applications and provisioning network resources accordingly, while the in reality the QoS requirements are time-varying.   
% TITAN addresses this mismatch by using higher-layer QoS profiles to design a dynamic SRA scheme for tactile-video traffic. More specifically, we design a slow-timescale resource allocation for 
% bandwidth slicing between tactile and video traffic and a fast-timescale scheduler for scheduling UEs on the allocated resources. 
Our performance evaluation demonstrates that TITAN improves joint QoS performance of tactile and video streams by \SI{7.64}{pp} over the best fixed-slicing scheme. 

% The results support dynamic QoS-aware SRA as a practical network-side
% complement to interaction-aware TI systems. The RAN need not infer the application-context  itself, but can respond effectively when time-varying QoS requirements are exposed through a QoS profile. In a future work, we will investigate
% online QoS-profile prediction and quantify TITAN's reduction of network control overhead along with further optimizations to TITAN to push its performance boundaries. 

% \section*{Acknowledgment}

% \vspace{-5pt}

\bibliographystyle{IEEEtran}
\bibliography{comprehensive_journal_abbr, reference_list}

\end{document}